\documentclass{osa-article}

\journal{osac}

\articletype{Research Article}

\usepackage{lineno}
\usepackage{subfigure}
\usepackage{algorithm}
\usepackage{algorithmic}
\renewcommand{\algorithmicrequire}{\textbf{Input:}}  
\renewcommand{\algorithmicensure}{\textbf{Output:}} 

\begin{document}
\title{Lensless coherent diffraction imaging based on spatial light modulator with unknown modulation curve}

\author{Hao Sha\authormark{1,4}, Chao He\authormark{2,4}, Shaowei Jiang \authormark{3},  Pengming Song\authormark{3}, Shuai Liu \authormark{2},  Wenzhen Zou \authormark{1}, Peiwu Qin \authormark{2}, Haoqian Wang \authormark{2},  and Yongbing Zhang\authormark{1,*}}

\address{\authormark{1}School of Computer Science and Technology, Harbin Institute of Technology (Shenzhen), Shenzhen, Guangdong 518006, China\\
\authormark{2}Tsinghua Shenzhen International Graduate School, Tsinghua University, Shenzhen, Guangdong 518006, China\\
\authormark{3}Department of Biomedical Engineering, University of Connecticut, Storrs, CT 06269, USA \\
\authormark{4}These authors contributed equally}

\email{\authormark{*}ybzhang08@hit.edu.cn} 



\begin{abstract}
Lensless imaging is a popular research field for the advantages of small size, wide field-of-view and low aberration in recent years. However, some traditional lensless imaging methods suffer from slow convergence, mechanical errors and conjugate solution interference, which limit its further application and development. In this work, we proposed a lensless imaging method based on spatial light modulator (SLM) with unknown modulation curve. In our imaging system, we use SLM to modulate the wavefront of object, and introduce the ptychographic scanning algorithm that is able to recover the complex amplitude information even the SLM modulation curve is inaccurate or unknown. In addition, we also design a split-beam interference experiment to calibrate the modulation curve of SLM, and using the calibrated modulation function as the initial value of the expended ptychography iterative engine  (ePIE) algorithm can improve the convergence speed. We further analyze the effect of modulation function, algorithm parameters and the characteristics of the coherent light source on the quality of reconstructed image. The simulated and real experiments show that the proposed method is superior to traditional mechanical scanning methods in terms of recovering speed and accuracy, with the recovering resolution up to 14 $\mu m$. 
\end{abstract}

\section{Introduction}

According to Huygens-Fresnel principle, diffraction occurs from each point in space. In lensless imaging system, the main task is to recover the optical field distribution of object from mixed signal without any optical lens for imaging or amplification \cite{Zhang2018OL}. Because lensless imaging system simplifies the illumination and imaging optical path and makes the overall system develop toward miniaturization and lightness, many research achievements have been made in the field of macro imaging \cite{Ozcan2008UltraWL, Zhang2017SP, Coskun2010LabChip,Zhang2020TIP}. In terms of microscopic imaging, conventional microscope system is unable to achieve both high resolution and wide field-of-view (FOV), and the spatial bandwidth product (SBP) is limited by the FOV, while lensless cameras can theoretically reach the diffraction limit resolution with the advantages of both high SBP and low aberration\cite{Ozcan2016ARBE,Guo2020JO}. 
In incoherent imaging system, the light intensity satisfies a linear relationship, so the image captured by sensor is the result of convolution of the object with the point spread function (PSF) of the system. The PSF can be calibrated by modulating the wavefront of the object with an encoding element (also known as mask). There are lots of researches on the design of mask in incoherent imaging system. R. Horisaki \emph{et al.} achieved single-shot phase imaging using amplitude mask with compressed sensing method \cite{Egami2016AO}. V. Ashok \emph{et al.} proposed to mount high-precision amplitude mask made by 3D printing onto sensors, enabling highly flexible macro and micro imaging \cite{Adamse2017SA}. T. Shimano \emph{et al.} used Fresnel diffraction grating as mask to calibrate the PSF \cite{Shimano18AO}. L. Waller \emph{et al.} utilized diffuser as the scatterer, which greatly reduced the complexity of calibration and was able to recover the 3D information of the object by calibrating at different depths \cite{Antipa2017Opt}. Although incoherent lensless modulation imaging is able to reconstruct the final image using deconvolution algorithms, mask-based reconstruction algorithm still suffers from insufficient luminous flux, poor frequency domain characteristics and slow convergence speed. In addition, incoherent imaging cannot recover the phase information of the object, limiting its further application.

Different from incoherent imaging, the complex amplitude is linear during the propagation of coherent light, making it possible to recover the phase information. Phase recovery can be divided into interference-based methods \cite{Brady09OE, Yair2018LSA, Wu2019LSA} and intensity-based methods \cite{Waller2010OE, Waller2010OE2}. The main application of the interference-based method is holographic imaging, where the principle is to convert phase information into observable intensity information by introducing a reference light that interferes with the light carrying object information \cite{Kim2006OP, Oh2010OE, Priyanka2017JBO}. Although the holographic imaging method requires a relatively low amount of data, it is also susceptible to the interference of conjugate solutions. Besides, the experiment is difficult to carry out due to the high requirements for the stability of the optical path. The intensity-based method applies phase recovery such as Gerchberg-Saxton (GS) algorithm or its variants for image reconstruction \cite{Song2012STQE, Nugent2010AP, Horisaki2016OE}. However, the GS algorithm requires objects to satisfy the sparsity condition and has relatively slow convergence. In contrast, ptychography iterative engine (PIE) algorithm solves the complex amplitude distribution of high-resolution samples by acquiring a series of low-resolution images \cite{Rodenburg2004APL}. G. Zheng \emph{et al.} placed the diffuser on the sensor surface and moved it through a two-dimensional high-precision displacement platform to modulate the object wavefront, enabling high-resolution pathology imaging with a wide field-of-view \cite{zheng2020LabChip}. C. Lu \emph{et al.} replaced mechanical scanning devices with LED arrays, to reduce the impact of mechanical errors\cite{Lu2021OE}. However, these methods rely on the movement of light sources or sensors, which will inevitably introduce the mechanical errors.

Another method of wavefront modulation is to use a spatial light modulator (SLM), which is capable of modulating light waves according to a given pattern. M. Deweert \emph{et al.} used SLM as a programmable amplitude mask to achieve incoherent imaging in natural light \cite{Deweet2015OEng}. Y. Wu \emph{et al.} achieved lensless high-resolution microscopic dynamic imaging of multiple samples and multiple scenes using high performance SLM \cite{Wu2019LSA2}. The reconstruction performance of these methods depends on the accuracy of the modulation pattern, which imposes a high demand on the SLM hardware.

To address these problems, in this paper, we report a lensless imaging method based on SLM and ptychographic algorithm. We employ a low-cost SLM to modulate the phase of the object wavefront and design an interferometric method to calibrate the modulation curve of the SLM. Combining with the ePIE algorithm, our method can recover the amplitude and phase information simultaneously, which improves the robustness and convergence speed of the system. The contributions of our work mainly include:

(1) We replace the mechanical displacement platform with a programmable random pattern and recover the amplitude and phase of the object utilizing ptychography scanning algorithm.

(2) We calibrate the modulation curve of low-cost SLM using split-beam interference experiment.

(3) We build a real coherent lensless imaging system and evaluate the effects of modulation function, algorithm parameters, and the light source on the quality of reconstructed image.

\section{Methods}

\subsection{Forward lensless imaging model}

\begin{figure}[htbp]
\centering\includegraphics[width=0.95\textwidth]{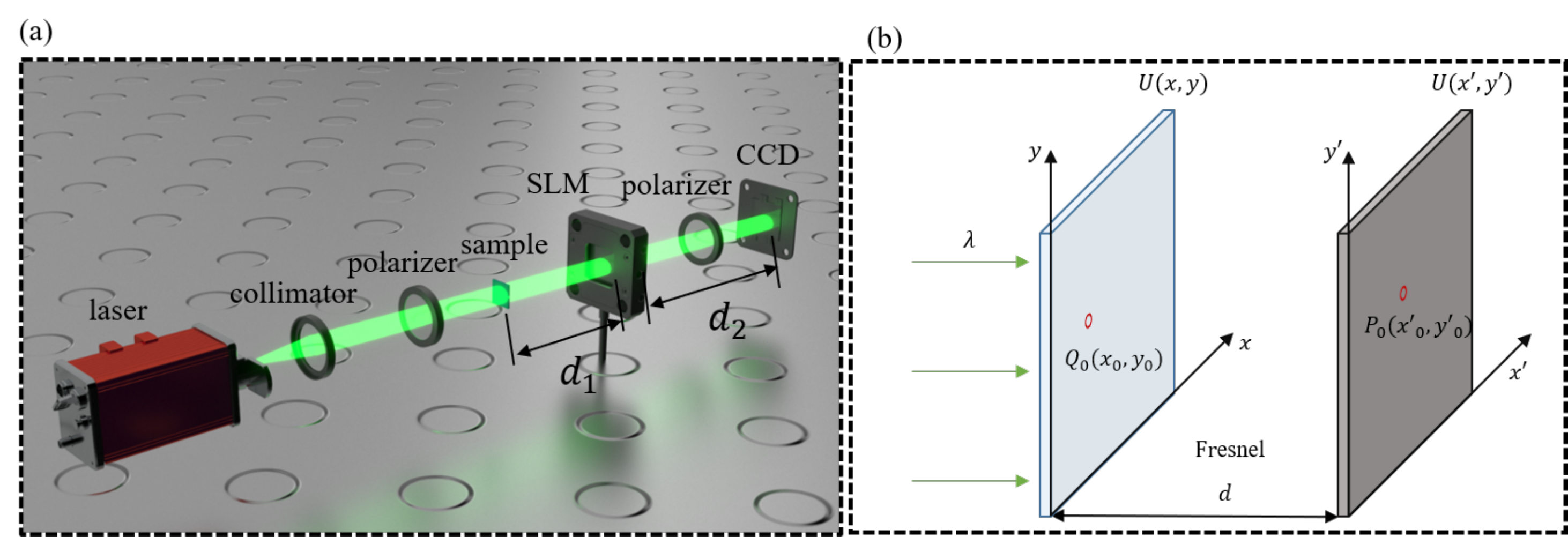}
\caption{The schematic diagram of proposed lensless imaging system. (a) The optical path of our system and (b) The Fresnel diffraction process.}
\label{Fig1}
\end{figure}

The schematic diagram of proposed lensless imaging system is shown in Fig.\ref{Fig1} (a). In this system, we utilize a laser as coherent light source and convert it to parallel light by a collimator. Inspired by \cite{Zhou2020OE}, we also insert two polarizers in the front and rear of the sample to fix the wavelength and polarization state of the light source to obtain stable modulation. The light of the object diffracts from the object plane to SLM plane after propagating $d_1$ distance. The SLM changes the complex amplitude state of the unit by controlling the directions of liquid crystal molecules through voltage, thus realizing the modulation. Finally, the modulated complex amplitude is propagated $d_2$ distance to the sensor plane. Since there is no lens in the imaging system, the object cannot be discerned from the image captured by CCD.

\begin{figure}[htbp]
\centering\includegraphics[width=0.95\textwidth]{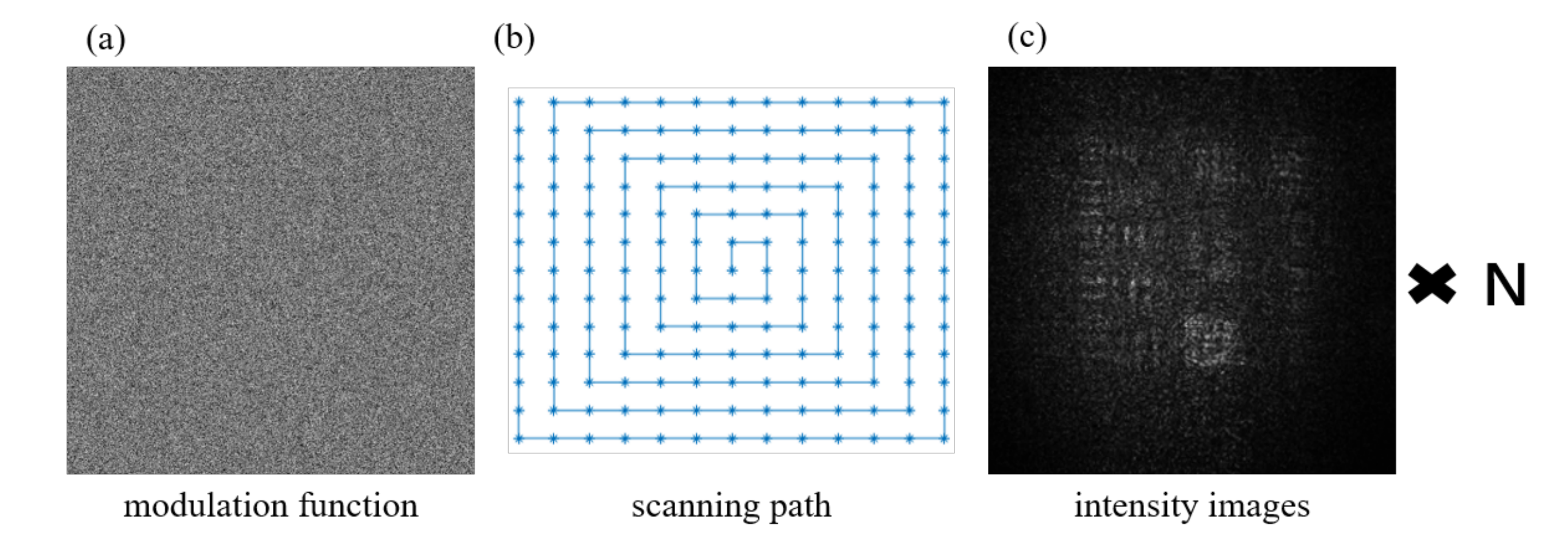}
\caption{The modulation process of SLM. (a)The random modulation function of the SLM, where the intensity of each pixel represents a complex value. (b) The scanning sequence of the pattern. (c) A series of diffraction intensity images of USAF resolution target captured by digital sensor under different patterns.}
\label{Fig2}
\end{figure}

A series of coherent diffraction images can be acquired by changing the modulation mode of the SLM. For the system shown in Fig.\ref{Fig1}(a), the intensity image produced by $n$-th modulated pattern on CCD can be summarized as:

\begin{equation}
I_n(\mathbf{r_{ccd}})=\left|prop_{d2}\left(SLM_n * prop_{d1}\left(O(\mathbf{r_{sample}})\right)\right)\right|^2,
\end{equation}  where $I_n(\mathbf{r_{ccd}})$ is the $n$-th intensity measurement on CCD, $O(\mathbf{r_{sample}})$ is the complex amplitude distribution of the object, $\mathbf{r}$ denotes the coordinates on the corresponding plane, $SLM_n$ is the modulation function, also named pattern, corresponding to the $n$-th measurement, $prop_d$ represents the diffraction process with propagation distance $d$. $d_{1}$ is the distance between object and SLM, and $d_{2}$ is the distance between SLM and CCD. As shown in Fig. \ref{Fig1}(b), when the diffraction distance is relatively small, the diffraction process can be approximated as Fresnel diffraction, and $prop_d$ has the following expression:

\begin{equation}
prop_{d}(O(x,y))=\frac{e^{\mathrm{j} k d}}{\mathrm{j} \lambda d} \iint_{-\infty}^{\infty} O(x, y) \exp \left\{\mathrm{j} \frac{k}{2 d}\left[(x-x')^{2}+(y-y')^{2}\right]\right\} d x d y,
\end{equation} where the $O(\cdot)$ is the complex amplitude distribution of the plane, $(x,y)$ and $(x',y')$ represent the coordinates of different diffraction planes respectively, $k=2\pi/\lambda$ is the wave vector, $\lambda$ is the wave length, and $d$ is the distance between the two diffraction planes. Further, we define an intermediate variable:

\begin{equation}
h(x, y)=\frac{e^ {\mathrm{j} k d}}{\mathrm{j} \lambda d}  \exp \left\{\mathrm{j} \frac{k}{2 d}\left(x^{2}+y^{2}\right)\right\} ,
\end{equation} then the Fresnel diffraction process $prop_d(O(x,y))$ can be expressed as :

\begin{equation}
prop_{d}(O(x,y))=O(x, y) \otimes h(x, y),
\end{equation}  where "$\otimes$" denotes convolution. Similarly, the image of the SLM plane can be calculated from the sensor plane according to the Fresnel inverse diffraction formula:

\begin{equation}
prop_{-d}(O(x',y'))=\frac{e^{\mathrm{j} k d}}{\mathrm{j} \lambda d} \iint_{-\infty}^{\infty} O(x', y') \exp \left\{-\mathrm{j} \frac{k}{2 d}\left[(x-x')^{2}+(y-y')^{2}\right]\right\} d x d y.
\end{equation} The modulation function $SLM_i$ is controlled by a grayscale image, where the intensity of each pixel represents a complex value $a+b\mathrm{j}$. The amplitude of this complex value represents the transmittance, ranged from 0 to 1, and its phase represents the phase modulation, ranged from $-1.2\pi$ to $1.2\pi$. To meet the requirements of the ptychography scanning algorithm, we use a random uniformly distributed image of size of $768\times768$ as a pattern, and a series of modulation functions with overlapping regions are generated by translating and stitching this pattern, as shown in Fig.\ref{Fig2} and \textbf{Visualization 1}.

\subsection{Image reconstruction algorithm}

The ePIE algorithm \cite{Maiden2009Utra} is not only able to reconstruct the unknown illumination function and the complex amplitude of object, but also has much higher robustness than traditional GS-based algorithms. In this paper, we employ the ePIE algorithm to jointly reconstruct the modulation function of SLM and the object function, which can greatly reduce the impact of SLM modulation errors. 

In this algorithm, the recovered object wavefront on SLM plane and the initial random uniformly pattern of the SLM for the first measurement are $O^{SLM}_0(\mathbf{r_{SLM}})$ and $P_0(\mathbf{r_{SLM}})$, respectively. Since there is only translational relations between different modulation function, the pattern of the $n$-th measurement can be expressed as $P_{n}\left(\mathbf{r_{SLM}}-\mathbf{s}_{\mathbf{n}}\right)$, where $\mathbf{s_n}$ denotes the pixel displacement between different patterns. The modulated complex amplitude $\Psi(\mathbf{r_{SLM}})$ is then propagated to the CCD plane by the Fresnel diffraction formula, and replace its amplitude with the square root of real image captured by digital sensor, $\sqrt{I_n(\mathbf{r_{ccd}})}$. The complex amplitudes after replacement in the CCD plane is further back propagated to the SLM plane, denoted by $\Psi'(\mathbf{r_{SLM}})$, according to the Fresnel inverse diffraction formula. The inputs of next iteration, $O^{SLM}_{n+1}(\mathbf{r_{SLM}})$ and $P_{n+1}(\mathbf{r_{SLM}})$, can be updated as:

\begin{equation}
O^{SLM}_{n+1}(\mathbf{r_{SLM}})=O^{SLM}_{n}(\mathbf{r_{SLM}})+\alpha \frac{\bar{P}_{n}\left(\mathbf{r_{SLM}}-\mathbf{s}_{n}\right)}{\left|P_{n}\left(\mathbf{r_{SLM}}-\mathbf{s}_{n}\right)\right|_{\max }^{2}}\left(\Psi_{n}^{\prime}(\mathbf{r_{SLM}})-\Psi_{n}(\mathbf{r_{SLM}})\right),
\end{equation}

\begin{equation}
P_{n+1}(\mathbf{r_{SLM}}-\mathbf{s}_{n})=P_{n}(\mathbf{r_{SLM}}-\mathbf{s}_{n})+\beta \frac{\bar{O}^{SLM}_{n}\left(\mathbf{r_{SLM}}\right)}{\left|O^{SLM}_{n}\left(\mathbf{r_{SLM}}\right)\right|_{\max }^{2}}\left(\Psi_{n}^{\prime}(\mathbf{r_{SLM}})-\Psi_{n}(\mathbf{r_{SLM}})\right),
\end{equation} where $\alpha$ and $\beta$ are the iterative update coefficients, usually taken as 1, and $\bar{P}_{n}$ and $\bar{O}^{SLM}_{n}$ denote the conjugate of the corresponding values respectively. The $P_{n+1}(\mathbf{r_{SLM}}-\mathbf{s}_{n})$ also needs to be re-shifted by $\mathbf{s}_{n}$ pixels to get the initial pattern, $P_{n+1}(\mathbf{r_{SLM}})$. To clearly demonstrate the ePIE-based lensless reconstruction algorithm, the pseudocode is given as below:

\begin{algorithm}
	\renewcommand{\algorithmicrequire}{\textbf{Input:}}
	\renewcommand{\algorithmicensure}{\textbf{Output:}}
	\caption{ePIE-based lensless reconstruction algorithm}
	\label{alg1}
	\begin{algorithmic}[1]
	    \REQUIRE Diffraction images $I_n(n=1,2,...,N)$ with the translational shift of the pattern
		\ENSURE Complex amplitude of the object $O(\mathbf{r_{sample}})$ and the modulation function of SLM $P(\mathbf{r_{SLM}})$
		
		\STATE Initialize the complex amplitude on SLM plane $O^{SLM}_{0}(\mathbf{r_{SLM}})$ and $P_{0}(\mathbf{r_{SLM}})$, and specify the number of iterations $M$
		\FOR{m=1:M}
		\FOR{n=1:N}
		\STATE $\Psi_{n}(\mathbf{r_{SLM}})=O^{SLM}_n(\mathbf{r_{SLM}})*P_n(\mathbf{r_{SLM}}-\mathbf{s_n})$
		\STATE $\Phi_{n}(\mathbf{r_{ccd}})=prop_{d2}(\Psi_{n}(\mathbf{r_{SLM}}))$
		\STATE $\Phi'_{n}(\mathbf{r_{ccd}})=\sqrt{I_n(\mathbf{r_{ccd}})} \exp{(j\cdot \angle{\Phi_{n}(\mathbf{r_{ccd}})})}$
		\STATE $\Psi'_{n}(\mathbf{r_{SLM}})=prop_{-d2}(\Phi'_{n}(\mathbf{r_{ccd}}))$
		\STATE Update the $O_{n+1}^{SLM}(\mathbf{r_{SLM}})$ using Eq. (6)
		\STATE Update the $P_{n+1}(\mathbf{r_{SLM}}-\mathbf{s}_{n})$ using Eq. (7)
		\STATE $P_{n+1}(\mathbf{r_{SLM}})$ = $P_{n+1}(\mathbf{r_{SLM}}-\mathbf{s}_{n} + \mathbf{s}_{n})$ 
		\ENDFOR
		\ENDFOR
	\STATE $O(\mathbf{r_{sample}})=prop_{-d1}(O^{SLM}_{N}(\mathbf{r_{SLM}}))$	
	\end{algorithmic}  
\end{algorithm}

\begin{figure}[htbp]
\centering\includegraphics[width=0.95\textwidth]{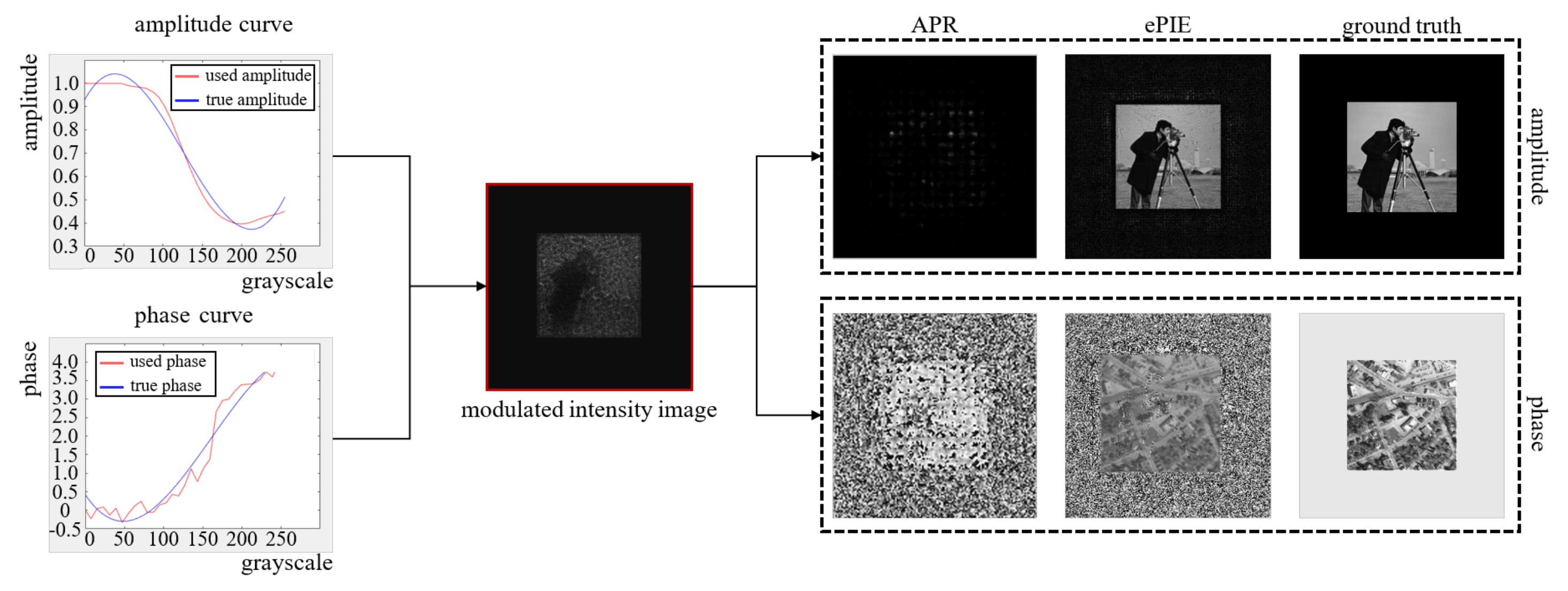}
\caption{The process of lensless imaging reconstruction based on ePIE algorithm.}
\label{Fig3}
\end{figure}

The simulation result is shown in Fig.\ref{Fig3}, where the red curve is the real SLM phase/amplitude-grayscale curve, and the blue one is the initial modulation function used for reconstruction. When there are errors between these two curves, the original image can be recovered by ePIE algorithm while traditional GS algorithms such as the Amplitude-phase retrieval (APR) algorithm cannot converge.

\subsection{SLM calibration}

Theoretically, the ePIE can recover the complex amplitude of the object even if the modulation function is unknown, but the selection of the initial value can greatly affect the convergence speed. Practically, the modulation function can be approximated from the grayscale image via the calibration curve, which is usually given at the factory. So, the modulation function of the SLM is usually taken as the initial pattern for ePIE. However, such SLM is usually more expensive, and the polarization state error also affects the accuracy of the calibration curve to some extent. Therefore, we approximately measure the amplitude-grayscale and the phase-grayscale curve through the optical power meter and light interference.

\begin{figure}[htbp]
\centering\includegraphics[width=0.95\textwidth]{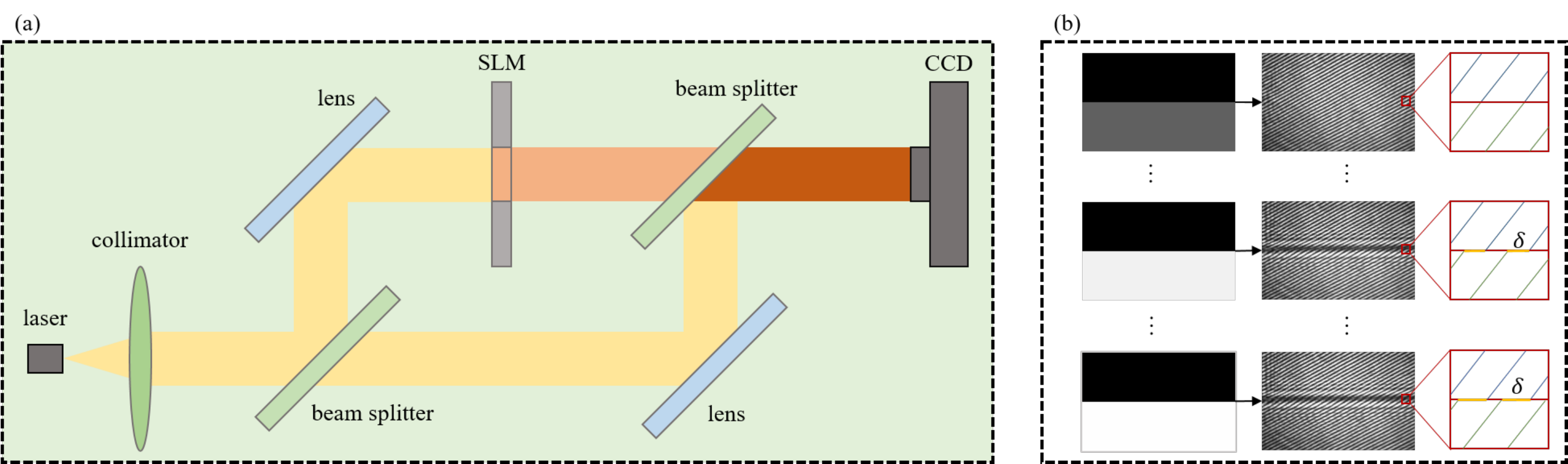}
\caption{The calibration experiment. (a) The optical path of split beam interference experiment. (b) The grayscale image and its corresponding interference fringe.}
\label{Fig4}
\end{figure}

The amplitude-grayscale calibration curve is relatively simple to measure by replacing the sensor with an optical power meter. In the measurement, the color range of [0-255] is scaled into [0,7,15,...,255], and the intensity corresponding to each grayscale image with a single value is recorded. The amplitude-grayscale calibration curve can thus be determined by interpolation. It should be noted that the phase-grayscale calibration curve records the phase difference under interference. The split beam interference optical path is shown in Fig.\ref{Fig4}(a). One beam of laser light reaches the sensor directly through the reflecting prism and the beam splitter, while the other beam passes through the SLM in its corresponding optical path. Since the SLM will modulate the complex amplitude of the light, there will be an optical path difference between two beams on the CCD plane, resulting in interference. The input grayscale image of SLM is shown in Fig.\ref{Fig4}(b), which consists of two different color blocks at the top and bottom. In the grayscale image, the upper half is filled with black as reference, and the lower varies from 0 to 255. When the grayscale value changes, the position of the interference fringe will also be shifted. By comparing the offset $\delta$ of the interference fringe corresponding to different grayscale values, the phase-grayscale calibrate curve is determined.

\section{Experiments and Results}

\subsection{Experimental set-up}

We build up a lensless coherent imaging system based on SLM for performance verification in real experiments. The laser generator (THORLABS S4FC520) generates a stable 520 nm coherent light source, which is then transformed into parallel light by a collimator. To verify the proposed method, we select a low-cost SLM (\$1500) with unknown modulation curve. The size of liquid crystal unit of SLM (tSLM-III) is 12.5 $\mu m$ and the number of pixel is 1024 × 768. After propagating a certain distance, the image sensor captures the modulated diffraction image. The image contains 2448 $\times$ 2048 pixels, and the pixel size is 3.45 $\mu m$. The hardware configuration for reconstruction is as follows: GPU of GeForce GTX Titan X, RAM of 32GB, and CPU of Intel(R) Core(TM) i9-9820X. The number of iterations is set to 10, and the reconstruction takes about 3 minutes.

In terms of modulation function, we use a noisy image conforming to a Gaussian distribution as the initial pattern and generate a sequence of 169 grayscale images based on a spiral scan with 2 pixels offset. The distance between the sample and the sensor ranges from 7 to 13 cm. 

\subsection{Reconstruction results}

\begin{figure}[htbp]
\centering\includegraphics[width=0.95\textwidth]{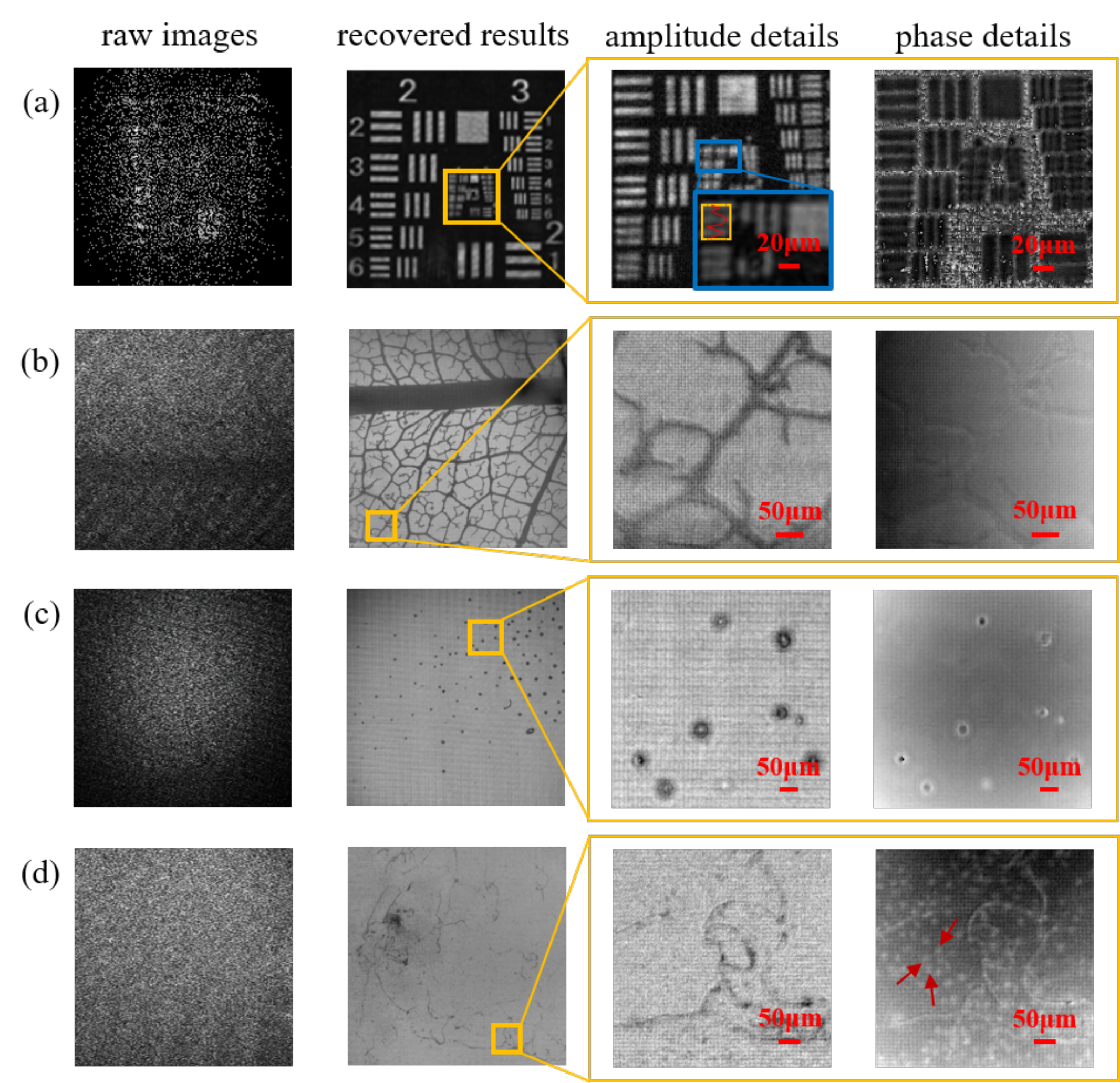}
\caption{The reconstructed results of the complex amplitude of (a) USAF-1951 resolution target, (b) leaf specimen, (c) stained cell slide, and (d) unstained cell slide. }
\label{Fig5}
\end{figure}

 In Fig.\ref{Fig5} (a), it is able to resolve (6,2) sets of the USAF target with resolution of 14 $\mu m$. Here (6,2) denotes the 6th group and the 2nd element in USAF, respectively, and the object is about 9 cm away from the sensor. The process of back propagation is also available in the \textbf{Visualization 2}. It can be seen that the resolution of our proposed system is very close to the size of liquid crystal unit of SLM. Similarly, in Fig.\ref{Fig5} (b), We can clearly distinguish the veins of the leave.

Fig.\ref{Fig5} (c) and (d) show the results of the reconstruction with and without staining the cells, respectively. In the amplitude image of Fig.\ref{Fig5} (c), the shape of the cell can be distinguished, and the dark region is the stained cytoplasm. While in the phase image, the nucleus is distinct for its higher refractive index. Further, we evaluated the recovery quality of unstained biological sample. As shown in Fig.\ref{Fig5} (d), the distribution of unstained cell is not visible in the amplitude image, but it can be clearly observed in the phase image (seeing red arrows in phase details). This reflects the characteristics of phase imaging, which enables imaging of samples with high transparency.

\begin{figure}[htbp]
\centering\includegraphics[width=0.85\textwidth]{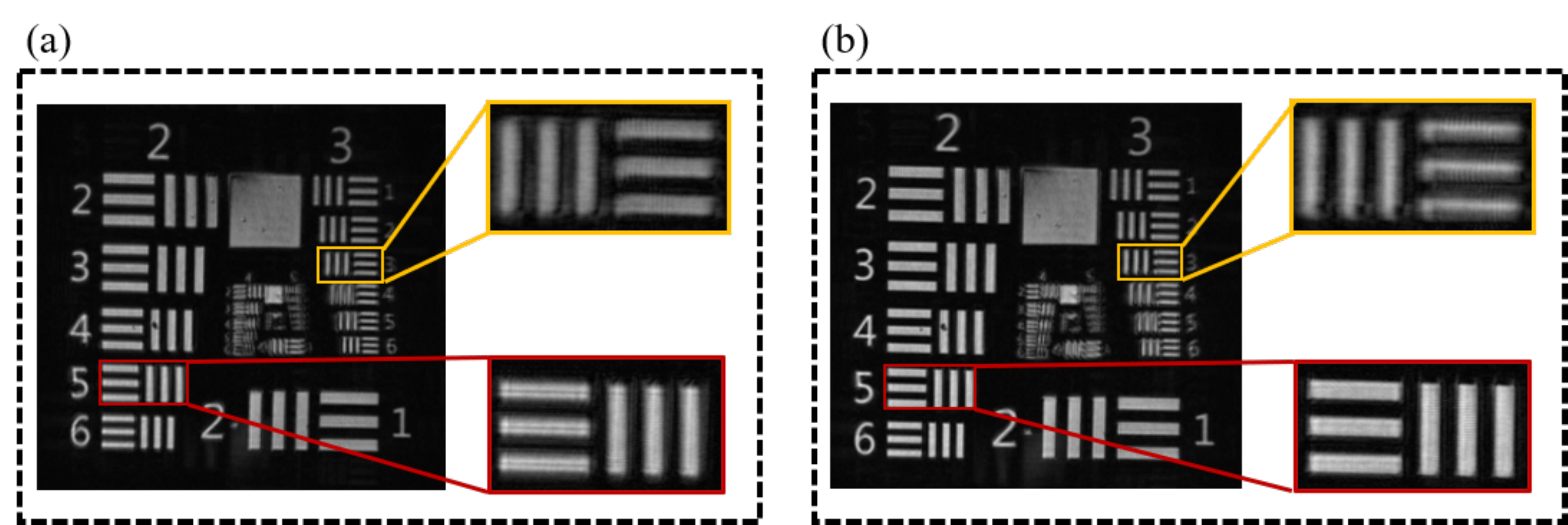}
\caption{Reconstruction results at different back-propagation distances: (a) back propagation 63 mm, and (b) back propagation 64.5 mm.}
\label{Fig6}
\end{figure}

Another application of lensless imaging is digital refocusing \cite{Li2018SP}. According to the diffraction equation Eq. (5), we can back-propagate the lensless imaging result to any depth and focus on a local region of the object. As shown in Fig.\ref{Fig6}, we conducted digital refocusing on samples tilted to the CCD plane, with the right side 1 cm farther from the sensor than the left side. Since the depth of the tilted USAF resolution target is very large, for a specific depth, the reconstructed image contains both in-focus and out-of-focus regions. For example, the regions within the red and yellow rectangles in Fig.\ref{Fig6} (a) correspond to the out-of-focus and in-focus, respectively. In contrast, the regions within the red and yellow rectangles in Fig.\ref{Fig6} (b) correspond to the in-focus and out-of-focus, respectively.

\subsection{Robustness analysis}

By incorporating the ePIE into our imaging system, our method can recover the complex amplitude of the object with high accuracy and fast speed, even if the modulation function is not exactly known. In this section, we discuss the effects of different factors on the reconstruction quality. 

\subsubsection{Effect of modulation function}

\begin{figure}[htbp]
\centering\includegraphics[width=0.65\textwidth]{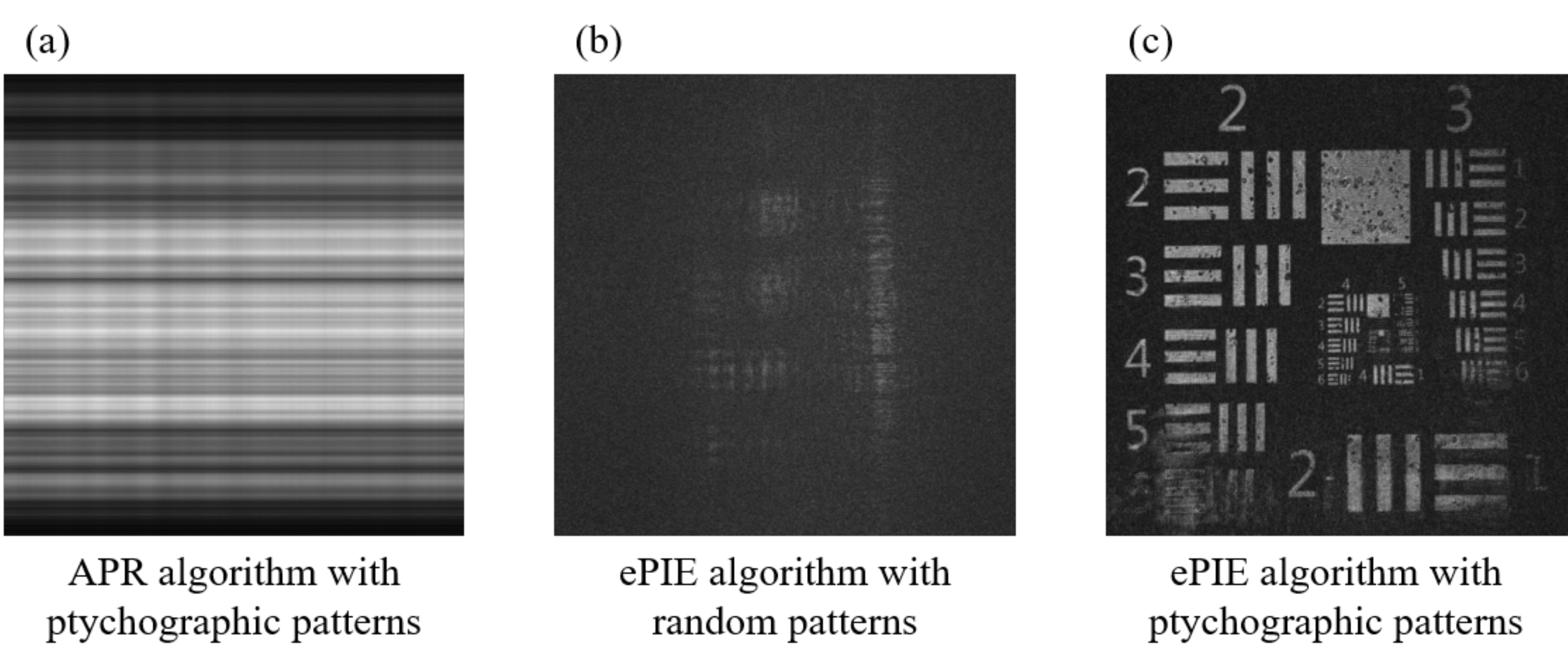}
\caption{ The reconstructed results using (a) spiral scan path and APR algorithm, (b) random scan path and ePIE algorithm, and (c) spiral scan path and ePIE algorithm.
}
\label{Fig7}
\end{figure}

In other SLM-based phase recovery algorithms, the calibrated curve is assumed to known and are generally reconstructed using 16 or 32 patterns that are independent of each other. In this paper, we use a series of patterns with a certain translation relationship for image reconstruction, and utlizing the ePIE algorithm can reduce the errors introduced by SLM itself. The effect of the relationship between different patterns on the reconstruction is shown in Fig.\ref{Fig7}. The reconstruction is performed using 32 intensity images in all experiments. It can be found only the combination of ptychographic patterns and ePIE algorithm can recover the information of the object, and when the SLM calibration curve has errors or using random patterns, the complex amplitude information cannot be reconstructed by either APR or ePIE algorithm.

\begin{figure}[htbp]
\centering\includegraphics[width=0.65\textwidth]{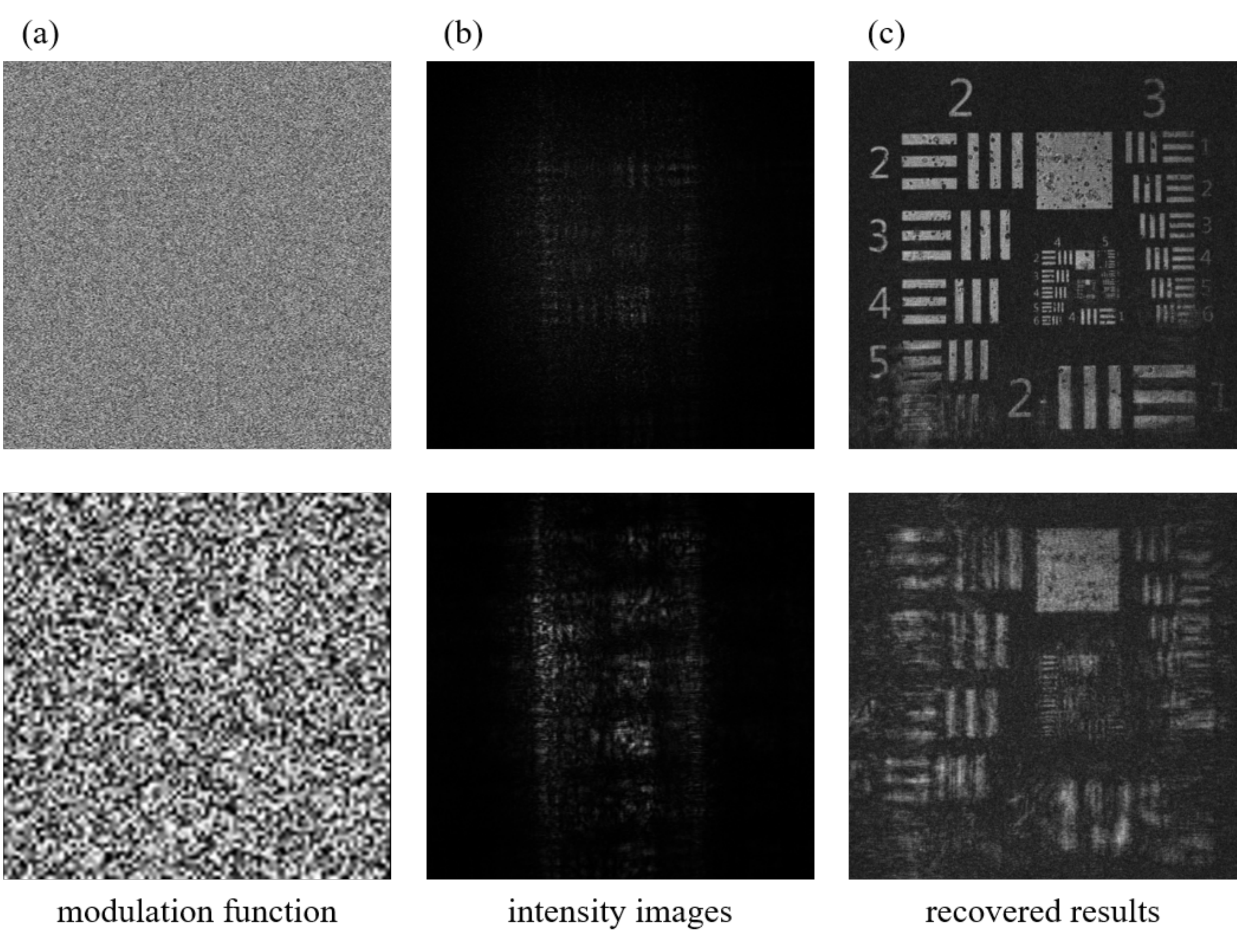}
\caption{The effect of the randomness of the modulation function on the reconstruction. (a) Different modulation function. The randomness of the upper modulation function is higher than the bottom. (b) The intensity image corresponding to the pattern. (c) The reconstructed results.}
\label{Fig8}
\end{figure}

It is stated the design of the pattern affects the reconstruction of the complex amplitude \cite{Cossairt2013TIP}, and we carried out an experiment to analyze the relationship between pattern modes and reconstruction quality. Since the resolution of SLM is 12.5 $\mu m$, which is much larger than the 3.45 $\mu m$ of the sensor, the randomness of the modulation function can greatly affect the reconstruction results, as shown in Fig.\ref{Fig8}. The bottom modulation function in Fig.\ref{Fig8} is obtained by upsampling a random noise image of size 96 $\times$ 96, and the upper one is satisfied with independent random distribution condition. It can be seen that the pattern with high randomness will have better reconstruction results.

\subsubsection{Effect of algorithm parameters}

\begin{figure}[htbp]
\centering\includegraphics[width=0.95\textwidth]{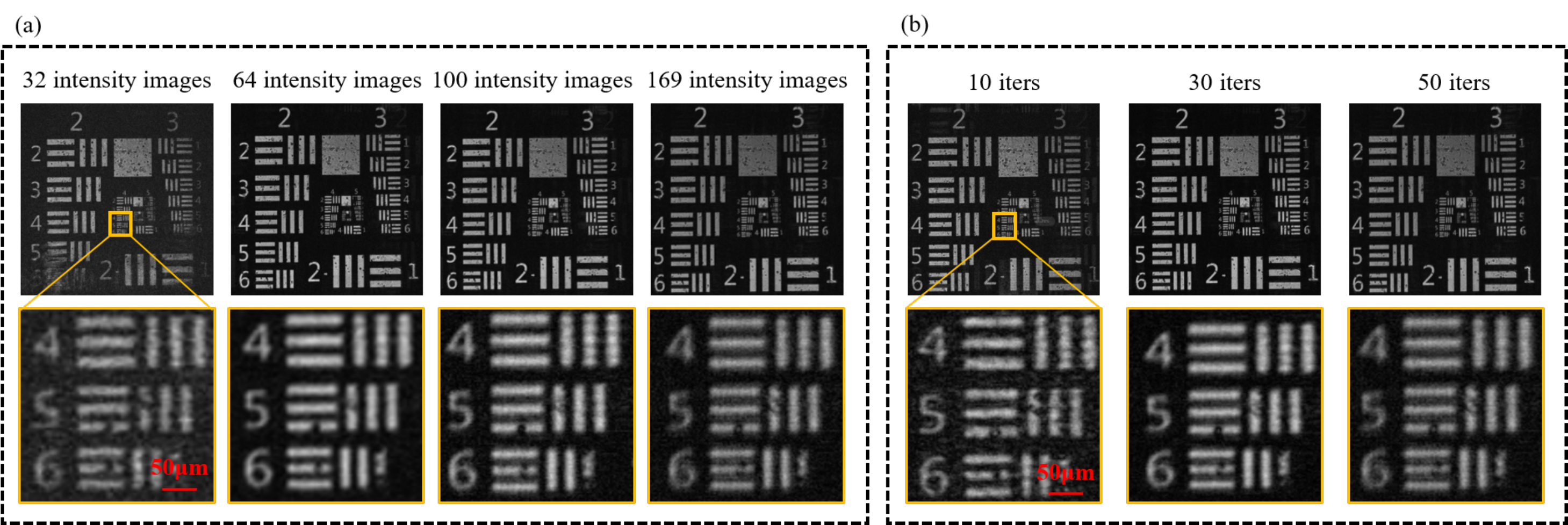}
\caption{The effect of the algorithm parameters on the reconstruction. (a) Reconstruction using different numbers of intensity images with 50 iterations. (b) Reconstruction using different numbers of iterations with 169 intensity images.}
\label{Fig9}
\end{figure}

The parameters that affect the reconstruction quality of ePIE algorithm mainly include the number of iterations and the number of captured intensity images. In Fig.\ref{Fig9} (a), we compared the effect of the number of intensity images on the reconstruction results. For USAF resolution target, the reconstruction quality does not get improved with the increase of captured intensity images, but stabilizes after a certain number. Besides, the fewer the intensity images used, the more iterations are needed. As shown in Fig.\ref{Fig9}(b), when 169 images are used, 30 iterations are sufficient to achieve the desired result, but 64 images require 50 iterations to achieve the corresponding result.

\begin{figure}[htbp]
\centering\includegraphics[width=0.65\textwidth]{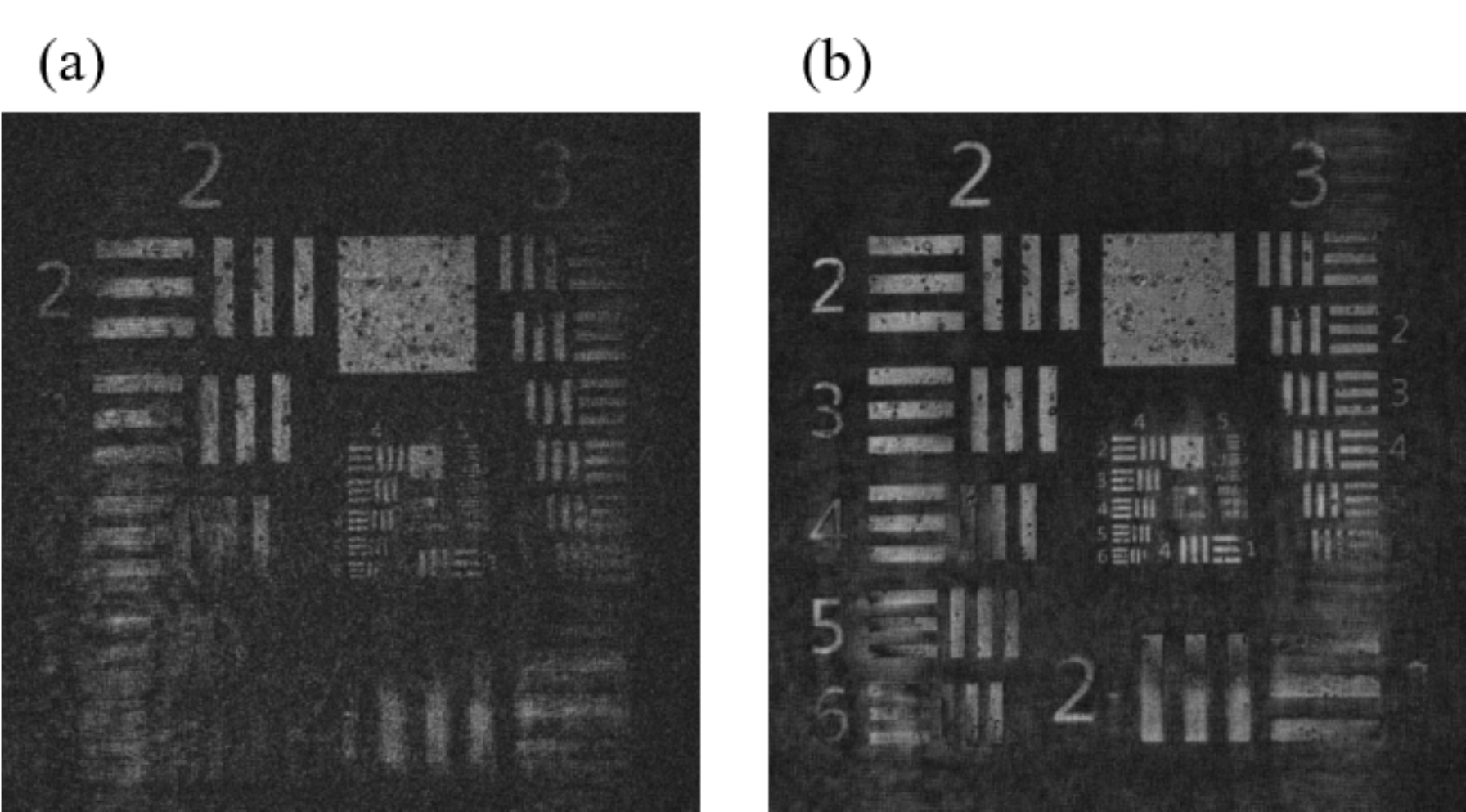}
\caption{The effect of the initial modulation function on the reconstruction.(a) Using random pattern as initial pattern. (b) Using calibrated pattern as initial pattern.}
\label{Fig10}
\end{figure}

On the other hand, the modulation curve determined by the calibration experiment can significantly improve the convergence speed of the algorithm, as shown in Fig.\ref{Fig10}. Fig.\ref{Fig10} (a) uses a random pattern as the initial pattern, while Fig.\ref{Fig10} (b) uses the modulated grayscale image of SLM as the initial pattern. In both Fig.\ref{Fig10} (a) and (b), 24 images are employed and 50 iterations are implemented. The result shows that using the calibrated pattern can reduce the interference of noise and improve the reconstruction quality. 

\subsubsection{Effect of light source}

\begin{figure}[htbp]
\centering\includegraphics[width=0.65\textwidth]{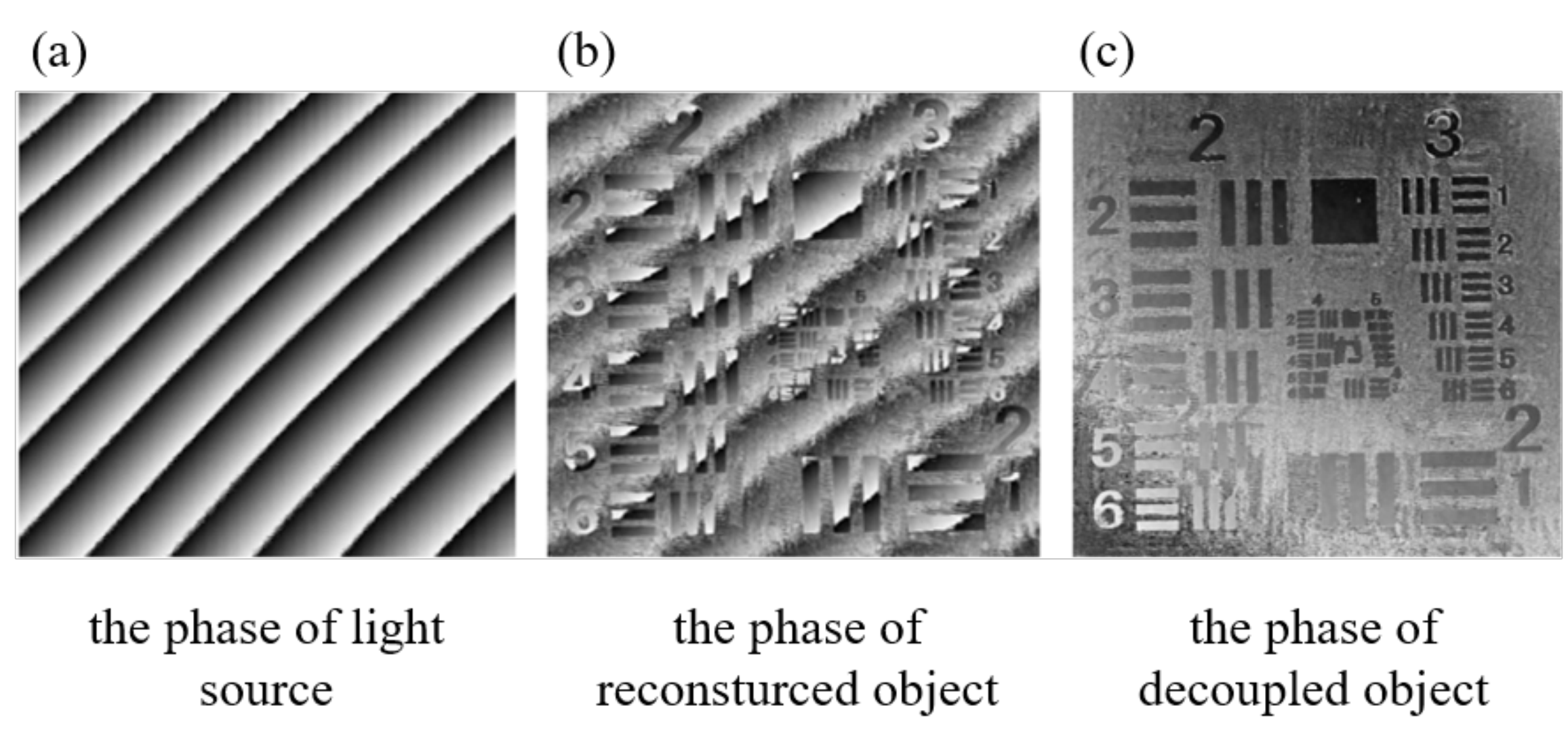}
\caption{Phase reconstruction results of (a) light source only, (b) object mixing with light source, and (c) decoupled object.}
\label{Fig11}
\end{figure}

Since the laser in our experiment is not an exactly parallel light source, it will interfere with the phase of the object in CCD plane, resulting in regular fringes in the phase image and affecting the final phase recovery quality. In this regard, we first measured the complex amplitude distribution of the light source with the same experimental parameters, and then put the sample into the optical path to eliminate the effect of the light source on phase recovery by phase filtering. As can be seen from Fig.\ref{Fig11} (a), the difference between the angle of  light incidence causes periodically varying tilted fringes in the phase image. These disturbances can couple with the complex amplitude of the object and affect the recovery quality, as shown in Fig.\ref{Fig11} (b). The decoupled phase image is shown in Fig.\ref{Fig11}(c), which significantly eliminates interference from laser sources.

\section{Discussion}

The experimental results have proven that our proposed method is able to reconstruct the complex amplitude information of the object without mechanical devices. We believe that the proposed method can be further applied in other tasks with the following factors considered:

(1) The acquisition rate and reconstructed resolution of the system in this study are still constrained by the sensor and the SLM itself. With high-precision hardware, the system has the potential to achieve higher resolution (close to 1 $\mu m$) and real-time dynamic imaging at ms level.

(2) The proposed lensless coherent imaging system can recover the wavefront of the object, and therefore has potential applications in the field of tomography. Conventional optical tomographic imaging often requires image acquisition at multiple focal planes to improve resolution. Our proposed system does not require any mechanical scanning and can achieve refocus at any depths according to the inverse propagation formula.

\section{Conclusion}

In this paper, we report a low aberration and wide FOV lensless imaging system based on SLM. First, we build a forward diffraction model of the object in coherent diffraction scenario. The simulation experimental results show that the object information can be recovered using the ePIE algorithm even the modulation function is not exactly known. Second, we design an optical interference experiment to coarsely calibrate the modulation curve of SLM, which can accelerate the convergence speed of the algorithm. Finally, we carry out a series of comparative experiments to evaluate the performance of our system. The reconstructed results show our lensless system can achieve to 14 $\mu m$ resolution and is able to reduce the requirement of the accuracy of SLM. In addition, we also discuss the effect of different parameters on the reconstruction quality. We believe this system can be further applied in other tasks such as tomography and microscopy in the future.

\begin{backmatter}
\bmsection{Funding}
This work was supported in part by the National Natural Science Foundation of China
under Grants 61922048 and 62031023, in part by the Shenzhen Science and Technology Project (JCYJ20200109142808034, JCYJ20180508152042002, and JSGG20191129110812708), and in part by Guangdong Special Support under Grant 2019TX05X187.

\bmsection{Disclosures}
The authors declare no conﬂicts of interest.

\bmsection{Data availability} Data underlying the results presented in this paper are not publicly available at this time but may be obtained from the authors upon reasonable request.

\end{backmatter}


\bibliography{sample}






\end{document}